\documentclass[epj]{svjour}

\usepackage{graphics}

\newcommand{\lsim}{\,{\buildrel < \over {_\sim}}\,}
\newcommand{\gsim}{\,{\buildrel > \over {_\sim}}\,}

\begin{document}
\title{Study of jet fragmentation in p+p collisions at 200 GeV in the STAR experiment.}

\author{Elena Bruna, for the STAR Collaboration
}

%
%
\institute{Physics Department, Yale University, 272 Whitney Avenue, New Haven CT-06520 (USA). elena.bruna@yale.edu}
\authorrunning{E. Bruna (STAR)}
\titlerunning{Study of jet fragmentation in p+p collisions at 200 GeV}
\date{Received: date / Revised version: date}


\abstract{
The measurement of jet fragmentation functions in p+p collisions at 200~GeV is of great interest because it provides a baseline to study jet 
quenching in heavy-ion collisions.
It is expected that jet quenching in nuclear matter modifies the jet energy and
 multiplicity distributions, as well as the jet hadrochemical composition. Therefore, a systematic study of 
the fragmentation functions for charged hadrons and identified particles 
is a goal both in p+p and Au+Au collisions at RHIC. Studying fragmentation functions for identified particles is interesting in p+p by itself because it provides a test of NLO calculations at RHIC energies.
We present a systematic comparison of jet energy spectra and fragment 
distributions using different jet-finding algorithms in p+p collisions in STAR.
Fragmentation functions of charged and neutral strange particles are also reported for different jet energies.
}
\maketitle

\section{Introduction}
RHIC experiments successfully reported the capability of high-$p_T$ particles to explore by parton energy loss the dense matter created in nuclear collisions~\cite{starhighpt,phenixhighpt}. However, such measurements are affected by well-known geometrical biases~\cite{renk} and high-momentum particles may come from low-energy jets that did not lose much energy in the medium. 
Therefore, there has been an effort to fully reconstruct jets in heavy-ion collisions in order to provide a direct measurement of the partonic kinematics, independently of the fragmentation process (quenched or unquenched).
The jet fragmentation function is described by the variable $\xi=\ln(E_{jet}/p_h)$, where $E_{jet}$ is the jet energy and $p_h$ is the momentum of the hadron in the jet. From an experimental point of view we rather use, in the absence of particle identification and at mid-rapidity, $\xi=\ln(p_{T,jet}/p_{T,h})$, where $p_{T,jet}$ is the reconstructed transverse momentum of the jet.
The $p_T$ spectrum of fully reconstructed jets in p+p collisions was measured by STAR with the 2003-2004 data, showing a good agreement with NLO calculations over the measured $p_T$ range up to 50 GeV~\cite{starjets1,starjets2}.
The increased statistics in the 2006 data-set due to the higher luminosity and detector acceptance allows one to reconstruct higher $p_T$ jets with better accuracy and to perform particle identification of jet fragments at high momentum.
Current analyses~\cite{joern,sevil} show that the measurement of reconstructed jets is feasible also in the high-multiplicity environment of Au+Au collisions at RHIC. First measurements of jet fragmentation functions in STAR were done both in p+p \cite{dong,mark} and Au+Au \cite{joern}.
In these proceedings we report a systematic study of fragmentation functions for charged particles in p+p for different jet finding algorithms. Given the inherent uncertainties in 
relating reconstructed jets to partons in QCD calculations and the 
complexity of jet-finding algorithms, it is essential to compare 
different algorithms to assess the stability of the procedures.
Fragmentation functions of neutral strange particles ($\Lambda$ and $K^0_s$) are also reported.
Measurements of fragmentation functions in p+p collisions are useful as a baseline for Au+Au data, in order to study the expected hadrochemical modification of jets in the medium~\cite{wiedemann}.

\section{Technical aspects}
\subsection{Event selection and technical setup for Jet-Finding}

This analysis is based on the p+p data at $\sqrt s$=200 GeV recorded by STAR during the 2005-2006 run at RHIC.
Two online triggers are used for high $p_T$ analyses in STAR: the High Tower (HT) and the Jet Patch (JP) trigger.
The HT trigger requires in addition to the minimum bias condition one tower in the Barrel Electromagnetic Calorimeter (BEMC) above a transverse-energy threshold of 5.4 GeV. The BEMC has full azimuthal ($\phi$) coverage and extends over pseudo-rapidity $-1<\eta<1$. Each tower has a size of $\Delta \eta \times \Delta \phi=0.05 \times 0.05$. 
In the JP trigger the total energy of a larger area of $\Delta \eta \times \Delta \phi=1 \times 1$ is required to be above a threshold of 8 GeV.
The total luminosities are 8.7 pb$^{-1}$ and 11 pb$^{-1}$ for JP and HT triggered events, respectively.
 
To avoid double-counting of electrons which leave signals both in the Time Projection Chamber (TPC) and in the BEMC,  we reject the BEMC energy that matches electron candidates and keep only the corresponding track $p_T$. In addition, since a charged hadron is expected to deposit part of its energy in the calorimeter, the energy of one MIP is subtracted from a tower hit by a charged hadron.

\subsection{Jet-Finding algorithms}\label{jetfind}
\begin{figure}
\centering
  \resizebox{0.45\textwidth}{!}{  \includegraphics{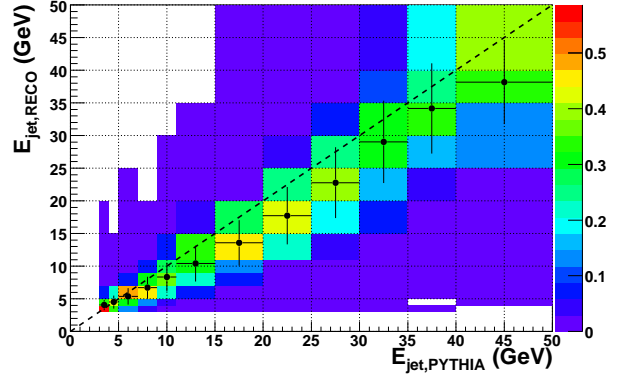}
}
  \caption{Comparison of RECO vs PYTHIA jets: the y-axis is the reconstructed jet energy after GEANT and track reconstruction; the x-axis is the energy of jets from PYTHIA generated particles. The vertical bars in each bin of $E_{jet,PYTHIA}$ are the RMS's of the histograms and represent the experimental energy resolution.}
  \label{fig:enres}
\end{figure}

A jet is experimentally defined as a cluster of particles in the $\eta-\phi$ plane. 
Jet-Finding algorithms can be grouped into two main categories: cone and recombination algorithms (see Ref.~\cite{reviewjets} for more details). 
Cone algorithms start from a ``seed'' above a given threshold, and look for other particles in a cone of radius R defined as  $R=\sqrt{(\Delta \phi^2 + \Delta \eta^2)}$. The seed is the energy of tracks or towers in a given bin in the $\eta-\phi$ plane. 
Cone-jet finders can be optimized by introducing the midpoint and the splitting/merging techniques~\cite{reviewjets} and iterations. When all the particles are used as seeds, the algorithm is called seedless. An example of that is the SISCone (Seedless Infrared Safe Cone) algorithm~\cite{SISc,SISc2}, which is part of the FastJet~\cite{fastjet} package, a C++ code for jet reconstruction.
Recombination algorithms are seedless by definition and are in general not bound to a circular structure. The clustering process is based on energy/distance weighted criteria with a distance scale R. The ``$k_T$'' algorithm starts from merging pairs of low-$p_T$ particles close in phase space. On the contrary, the ``anti-$k_T$'' starts with high momentum particles~\cite{antikt}. Both ``$k_T$'' and ``anti-$k_T$'' are part of the FastJet package.
The jet energy resolution can be estimated with a PYTHIA simulation of p+p collisions at $\sqrt s$=200 GeV. The jet finder is applied both to the PYTHIA output and to the output of the GEANT detector simulation plus the standard reconstruction chain. Fig.~\ref{fig:enres} shows the comparison between the energies of jets made of Monte Carlo particles (PYTHIA jets) and those built of reconstructed particles (RECO jets). The jet algorithm is a midpoint cone algorithm with R=0.7, a seed of $E_{seed}=0.5$ GeV, a minimum-$p_T$ cut-off of 0.1 GeV/c and a splitting/merging fraction $f_{split}=0.5$. The energy resolution is $\sim 25\%$ and is approximately constant for all the measured energies. 
As reported in Fig.~\ref{fig:enres}, the jet-energy obtained with reconstructed particles is on the average smaller than the energy of the PYTHIA-jet: this is due to the missing energy of neutral particles like neutrons and $K^0_L$ in the reconstructed jets.
An alternative way to estimate the jet-energy resolution is to study di-jets, and compare the energies of the two reconstructed back-to-back  jets. We found that the energy resolution calculated with this method is in agreement with that obtained with simulation, indicating that $k_T$-broadening is 
small compared to the jet-energy. The di-jet analysis is promising because it offers a complete model independent way to study jets.
The results shown in the following refer to jets with the highest energy per event.

\section{Reults}
\subsection{Jet $p_T$ spectra}
\begin{figure}
\centering
\resizebox{0.4\textwidth}{!}{  \includegraphics{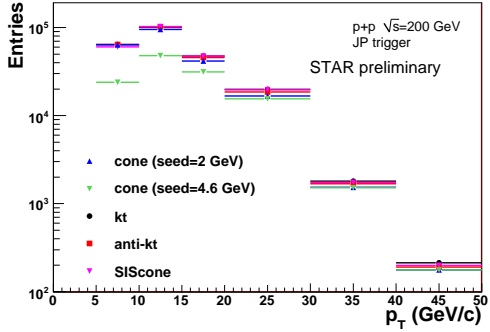}
}
  \caption{Uncorrected reconstructed $p_T$-spectrum of jets for different jet finders: cone with $E_{seed}=2$ GeV (blue), cone with $E_{seed}=4.6$ GeV (green), $k_T$ (black), anti-$k_T$ (red), seedless SIS cone (purple).}
  \label{fig:jetpt}
\end{figure}

We ran four different jet finders on the JP triggered events in order to compare their performance. The algorithms, described in Section~\ref{jetfind}, are: (1) Leading-Order High-Seed Cone (LOHSCone~\cite{joern,sevil}) without any additional option like splitting/merging and two choices of seeds (2 and 4.6 GeV), (2) $k_T$, (3) anti-$k_T$ and (4) seedless SIS cone. 
Fig.~\ref{fig:jetpt} shows the jet $p_T$ spectra for the four jet finders with R=0.7. The jets are required to be entirely in the detector acceptances, therefore a cut $|\eta_{jet}|<0.3$ is applied ($\eta_{jet}$ is the pseudo-rapidity of the centroid of the jet). 
No correction for jet energy resolution and trigger bias was applied.
Bin widths are chosen to be compatible with the experimental jet-energy resolution of $\sim25\%$. All the FastJet algorithms plus the cone with the lower seed ($E_{seed}=2$ GeV) agree with each other, suggesting that all of them can be employed for jet analyses in p+p.
The cone jet-finder with a high seed shows a lower 
yield at lower jet-energy because it selects jets with a high-z leading 
fragment ($z=p_{T,h}/p_{T,jet}$). This effect increases with the seed.

\subsection{Jet Fragmentation Functions}
\subsubsection{Assessing trigger biases}
\begin{figure}[h]
\centering
\resizebox{0.4\textwidth}{!}{  \includegraphics{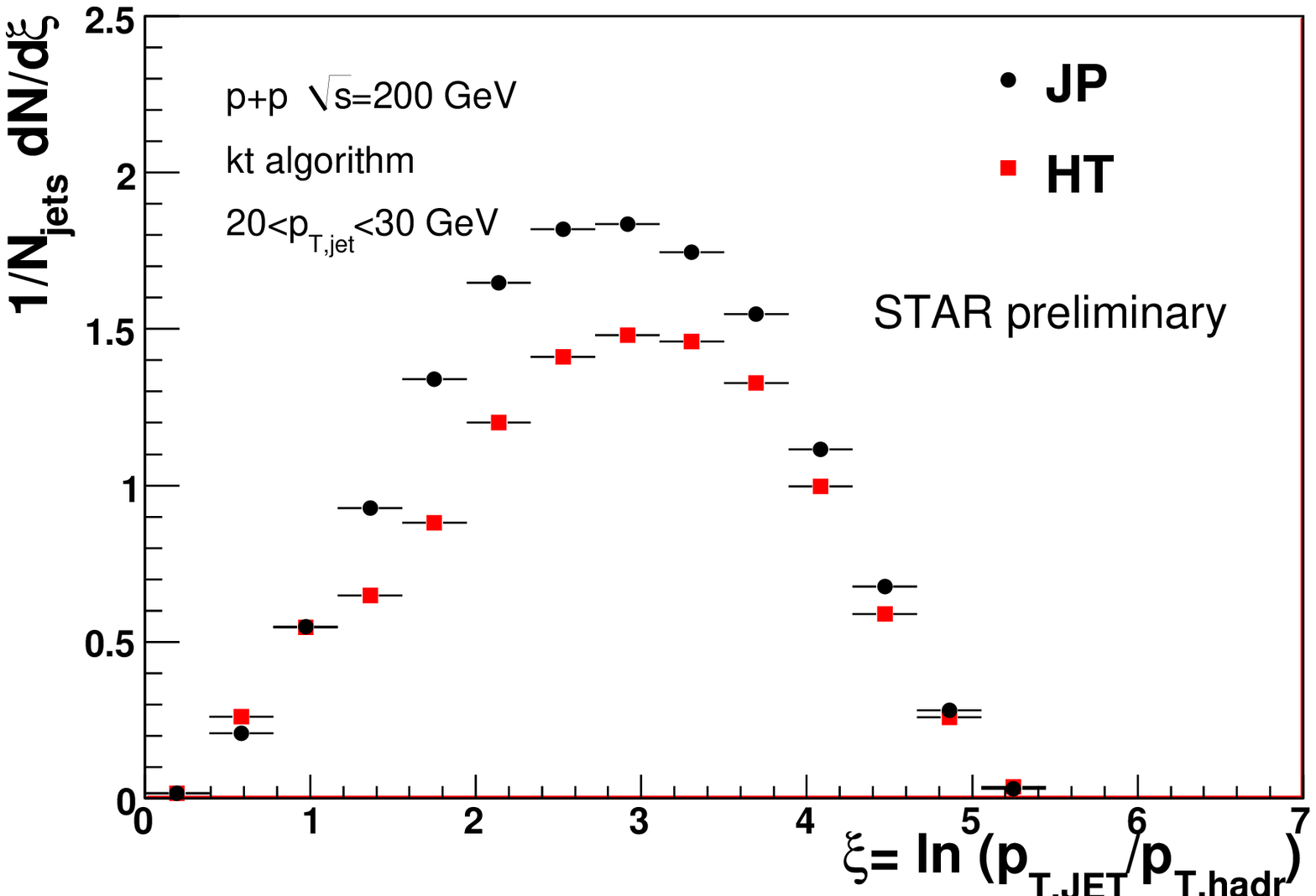}}
\resizebox{0.4\textwidth}{!}{  \includegraphics{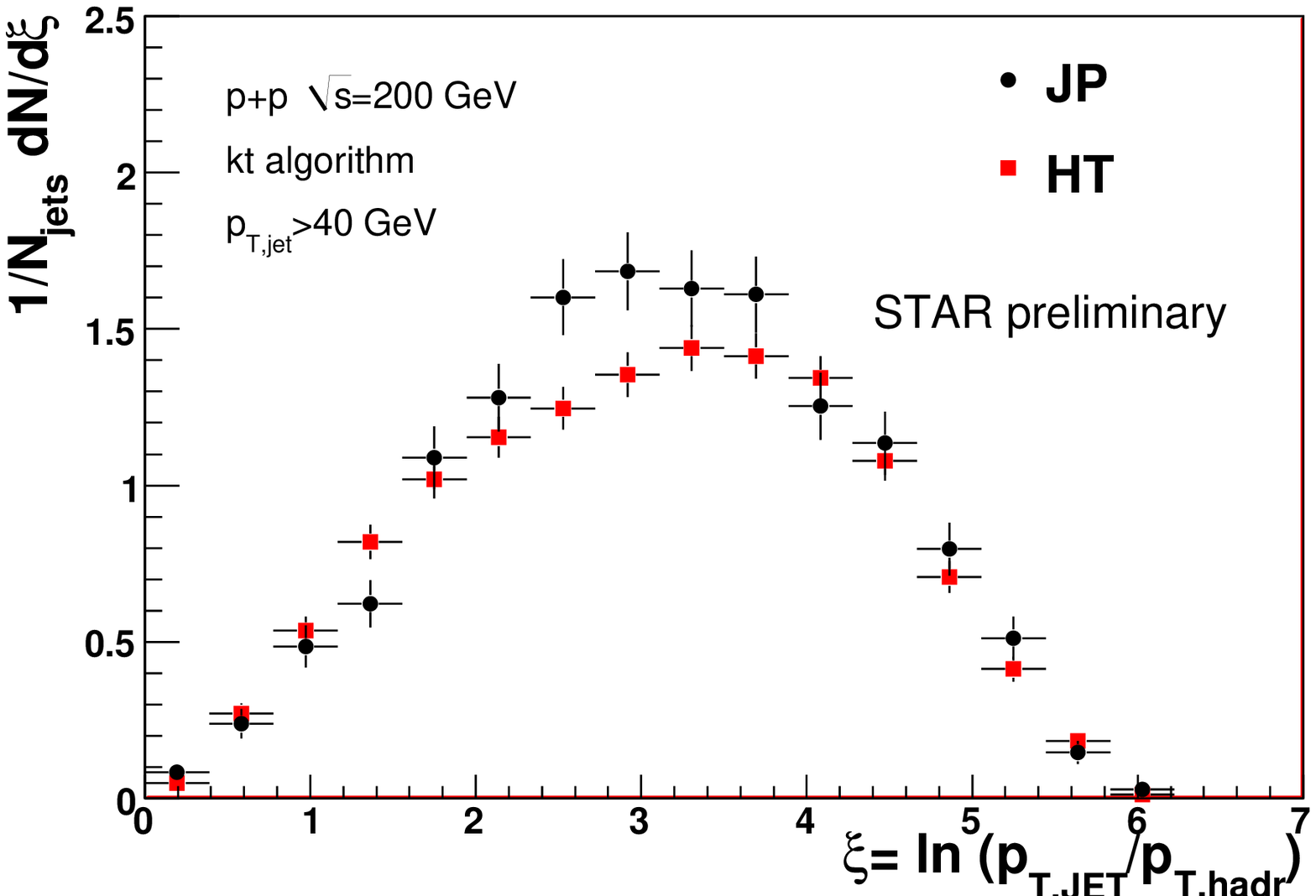}}
\caption{Upper plot: uncorrected $\xi$ distribution for JP (red) and HT (black) triggered data for $20<p_{T,jet}<30$ GeV. The results are shown for the $k_T$ algorithm with R=0.7. Bottom plot: the same as upper plot but for  $p_{T,jet}>40$ GeV.}
\label{fig:JPHT}
\end{figure}

In order to assess the biases in the two online trigger selections, the jet fragmentation functions for HT and JP data are compared for different jet energies (a similar study was done before using the neutral energy fraction of HT and JP jets~\cite{starjets2}). The $\xi$ distributions of charged particles are normalized to the number of jets, so that the integral of the spectrum is the average multiplicity of the charged particles in a jet of a given energy. The fragmentation functions are shown in Fig.~\ref{fig:JPHT} for JP (black circles) and HT (red squares) for two jet energies, $20<p_{T,jet}<30$ GeV and $p_{T,jet}>40$ GeV respectively. The bin width is compatible with the jet energy resolution.

At jet energies below $\sim30$ GeV the multiplicity of charged particles in jets is significantly different for HT and JP data. 
The observed difference is consistent with a stronger trigger bias in HT 
events: the HT trigger selects jets with a leading neutral fragment 
carrying a rather large fraction of the jet-energy, thus depleting the charged hadrons at large $z$.
This results in a smaller multiplicity of charged particles for HT jets than that of JP jets (Fig.~\ref{fig:JPHT}, upper panel). This effect is reduced at higher jet energies, since the bias is less dominant (Fig.~\ref{fig:JPHT}, lower panel). Both JP and HT triggers are expected to show biases at low $\xi$ (high hadron $p_T$) which diminish for higher jet $p_T$. JP jets exhibit a negligible bias for $p_{T,jet}\gsim 20$ GeV~\cite{starjets2}.
\subsubsection{Comparison of Jet Algorithms}

\begin{figure}
\centering
\resizebox{0.45\textwidth}{!}{  \includegraphics{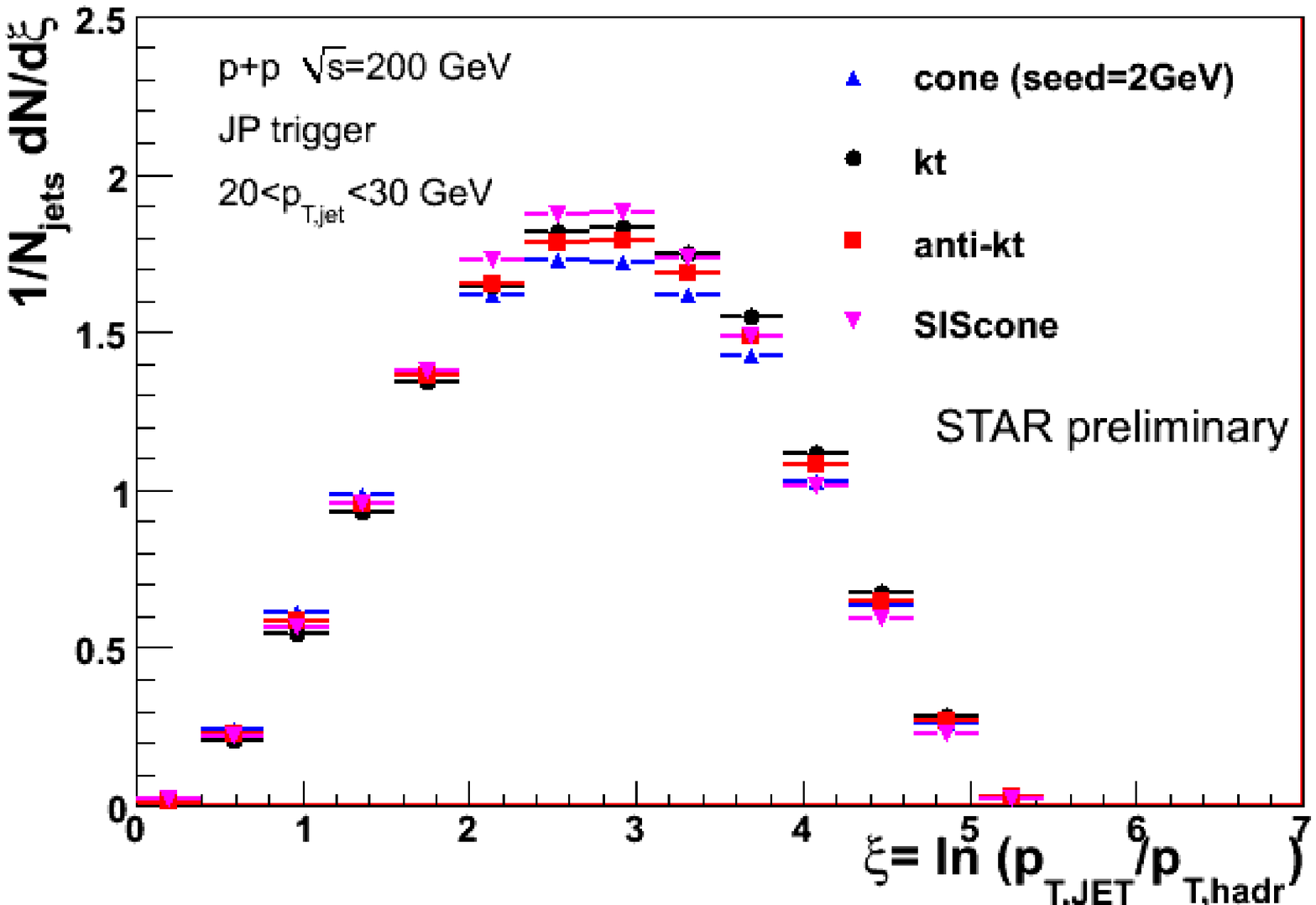}}
\resizebox{0.45\textwidth}{!}{  \includegraphics{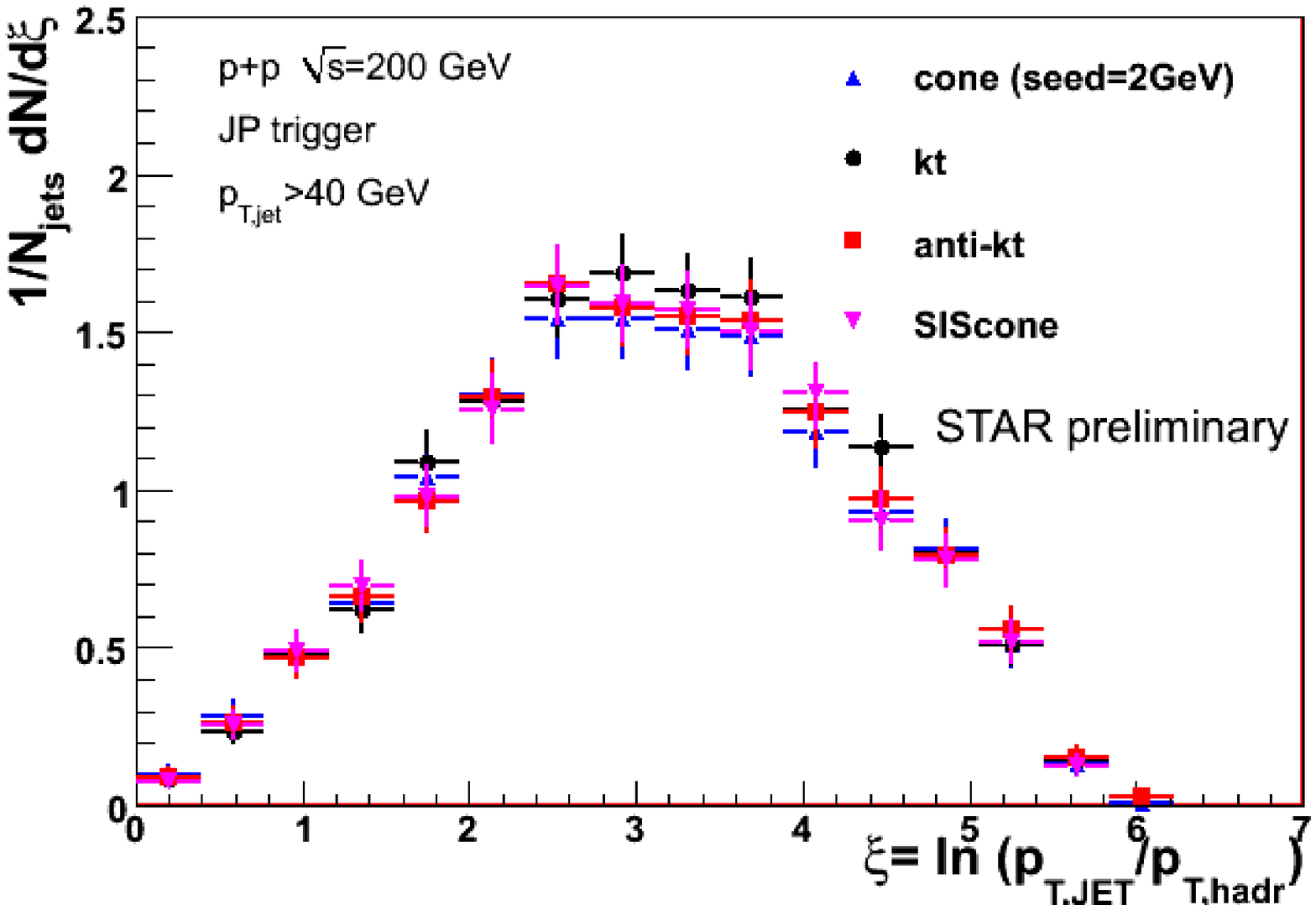}}

\caption{Upper plot: uncorrected $\xi$ distributions for the four different jet algorithms for $20<p_{T,jet}<30$ GeV with JP data and R=0.7. Bottom plot: the same as upper plot but for  $p_{T,jet}>40$ GeV.}
\label{fig:FF}
\end{figure}

The comparison of the four jet finders at the level of fragmentation functions is reported in Fig.~\ref{fig:FF} for JP data and R=0.7. All the jet algorithms agree over the range of jet energies explored, from 10 GeV to above 40 GeV, therefore all of them can be used interchangeably for jet analyses in p+p. A similar study was also performed with R=0.4 (radius used in jet reconstruction in Au+Au) with similar performance of the jet algorithms. 
This suggests that at RHIC energies NLO effects like soft gluon radiation are minor, otherwise cone and recombination algorithms could behave differently especially in the case  of smaller jet radii.


\section{Fragmentation functions of identified particles}
\begin{figure}
\centering
 \resizebox{0.3\textwidth}{!}{ \includegraphics{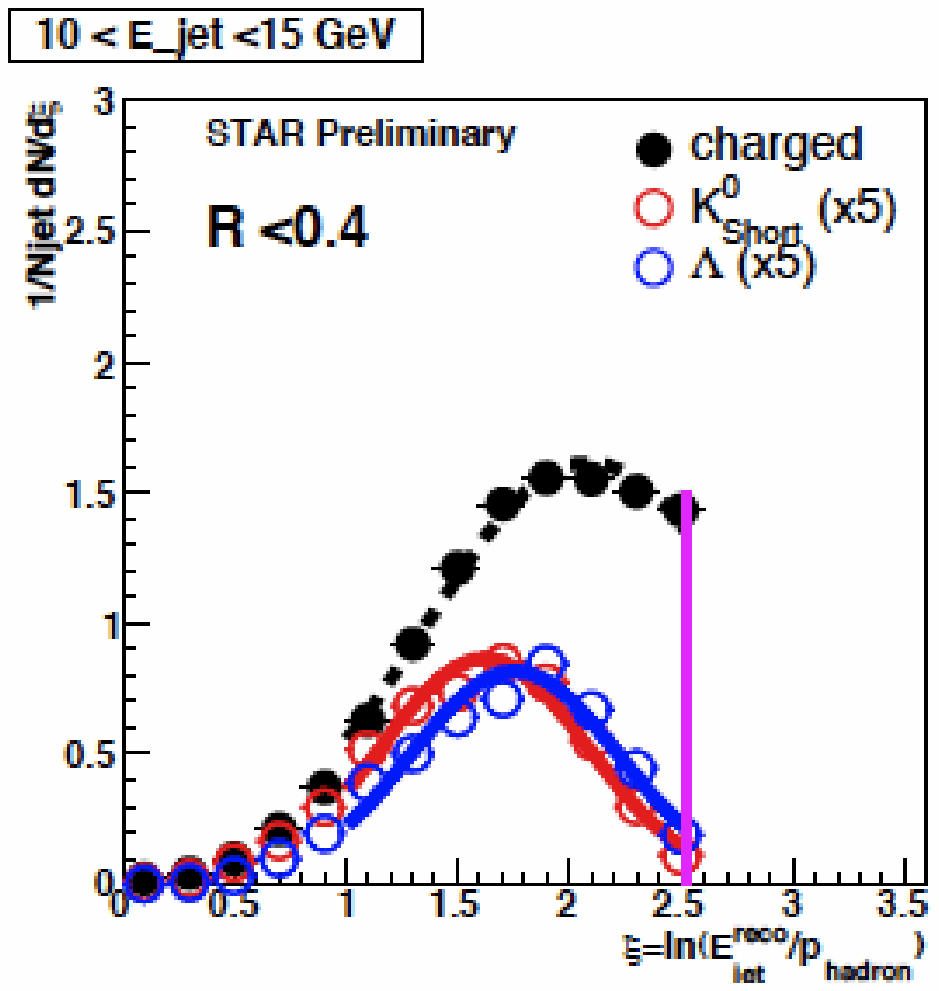}}
 \resizebox{0.3\textwidth}{!}{ \includegraphics{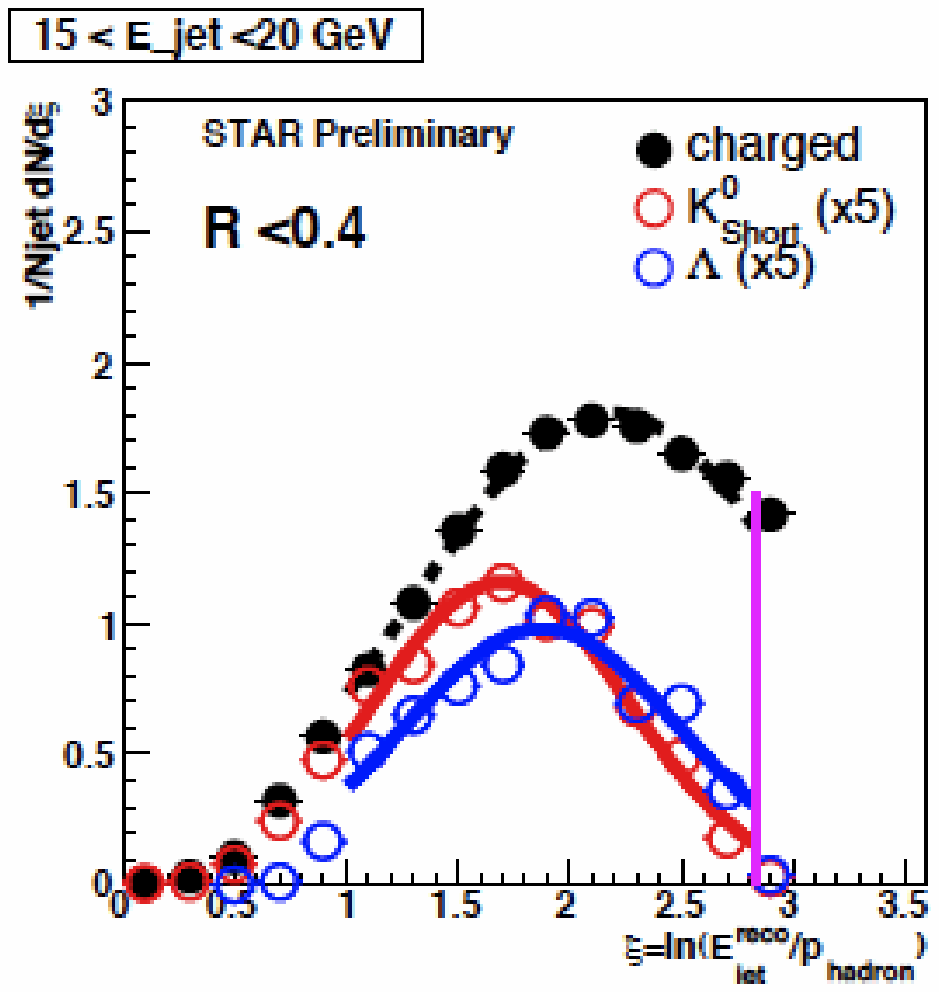}}
 \resizebox{0.3\textwidth}{!}{ \includegraphics{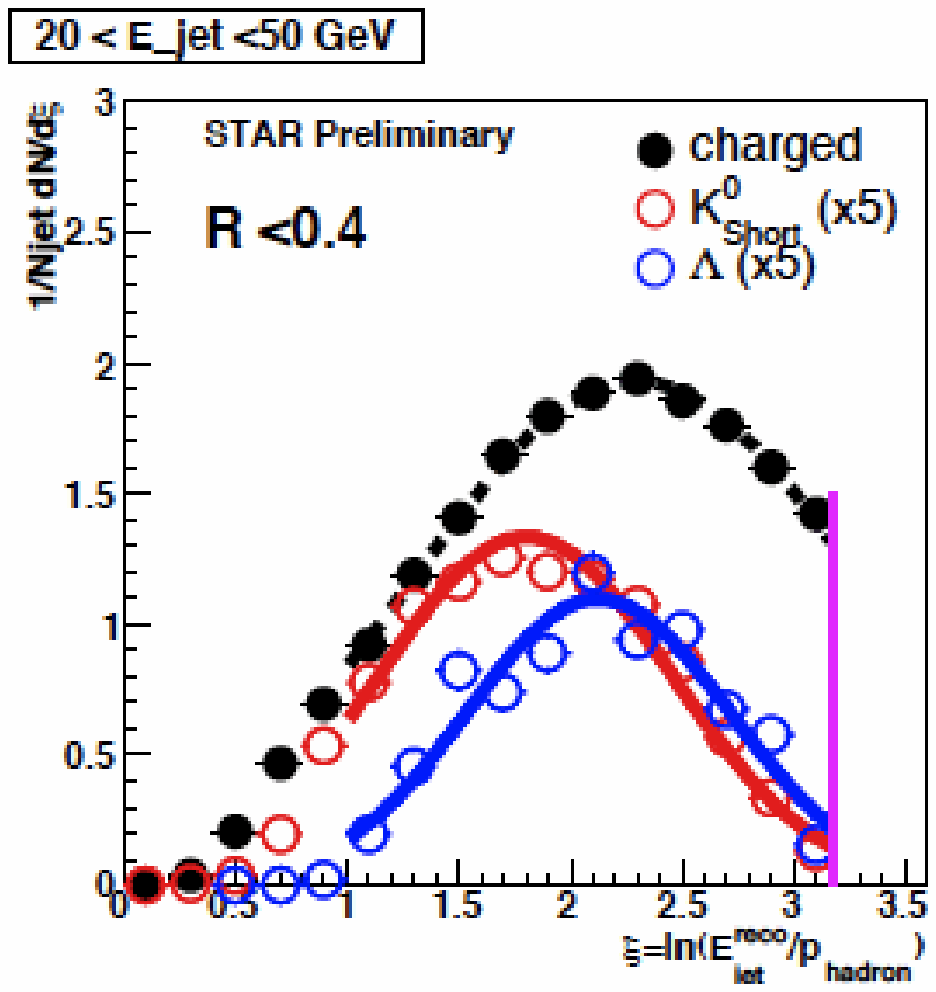}}
  \caption{$\xi$ distributions for strange particles in jets for three different jet energies. Gaussian fits are used to obtain the peak value $\xi_0$. The purple line indicates the $p_T$-cutoff of 1 GeV for good $K^0_s/\Lambda$ identification. Charged hadrons are shown for comparison and have a $p_T$-cutoff of 0.5 GeV. Taken from~\cite{mark}.}
  \label{fig:mark}
\end{figure}

STAR has excellent efficiency in finding V0 particles over a large kinematical range ($1\lsim p_T \lsim 8$ GeV/c), which allows to study jet fragmentation for identified particles, namely $\Lambda$ and $K^0_s$~\cite{mark}.
Once the V0's are reconstructed via invariant mass analyses, their decay particles are replaced by the V0 particle before the jet-finding step. The jet algorithm is a midpoint-cone with splitting/merging, $E_{seed}=0.5$ GeV, R=0.4 and $f_{split}=0.5$. Fig.~\ref{fig:mark} shows $\xi$ distributions for identified $\Lambda$ and $K^0_s$ for three jet energies.
QCD models predict a decrease of the peak value, $\xi_0$, with the mass. It is found that $\xi_0$ of strange particles is smaller than that of charged particles (mainly pions). However, there are indications of an inverse mass ordering of $\xi_0$ for strange particles because the $\xi_0$ of $\Lambda$ is larger than that of $K^0_s$. Detailed systematic uncertainities associated with these measurements are under investigation.
Similar observations with kaons and protons were reported by other experiments over a large energy range~\cite{anulli,trento}. 
Ongoing studies are focused on the comparison of fragmentation functions of strange particles with NLO calculations, and on the comparison of $p_T$ spectra of strange particles in jets and in the underlying event~\cite{star-strange1,star-strange2}.

\section{Summary and Outlook}
In these proceedings we report on a STAR measurement of uncorrected jet-$p_T$ spectra up to ~50 GeV and jet fragmentation functions for different jet energies in p+p at $\sqrt s$=200 GeV.
Two trigger selections were compared, and it was shown that the High Tower and the Jet Patch triggers agree in the fragmentation functions at high jet energies, $p_{T,jet}>40$ GeV, where the trigger bias is believed to be gone.
Despite the apparent large differences between 
different jet-finding algorithms, the measured energy spectra and 
fragmentation functions are very similar for different jet radii, suggesting that all of the algorithms can be interchangeably chosen in jet analyses in p+p. This is an important result considering p+p as baseline for jet reconstruction in Au+Au. In addition, the comparison of different algorithms gives access to systematic errors related to the jet finding procedure.
In the future the identification of jet fragments already started with strange particles will be pursued and extended to high-momentum particles, exploiting the relativistic rise. The goal is to compare fragmentation functions of identified particles in jets in p+p and Au+Au and study the expected increase of hadrochemical ratios like $K/\pi$ and $p/\pi$ at high-$p_T$ due to medium modifications: this would reflect in an increased difference between the distributions inside the jet and in the underlying event.
Analyses of di-jets will play a significant role in jet studies because they provide information on jet-energy resolution in a complete model independent way.

\end{document}